\documentclass[3p,10pt,a4paper,twoside,fleqn,sort&compress]{elsarticle}
\makeatletter
\def\ps@pprintTitle{%
  \let\@oddhead\@empty
  \let\@evenhead\@empty
  \let\@oddfoot\@empty
  \let\@evenfoot\@oddfoot
} \makeatother

\usepackage{graphicx}
\usepackage{verbatim}
\usepackage{amsmath}
\usepackage[varg]{pxfonts}
\usepackage{eepic}
\usepackage{amssymb}

\usepackage{mathtools}
\usepackage[all,cmtip]{xy}

\usepackage{tikz}
\usetikzlibrary{matrix,arrows}

\newtheorem{theorem}{Theorem}[section]

\newtheorem{remark}[theorem]{Remark}

\newcommand{\mc}{\mathcal}
\newcommand{\comp}{\circ}

\newcommand{\mcP}{P}
\newcommand{\mcU}{U}
\newcommand{\mcQ}{Q}
\newcommand{\mcPQ}{PQ}

\newcommand{\Hv}{\mc N}

\newcommand{\isoXtoR}{h^{-1}}
\newcommand{\isoRtoX}{h}

\newcommand{\Logp}{\hbox{Log}_q}

\newcommand{\op}{\oplus}
\newcommand{\cop}{\tau}

\newcommand{\ra}{\rightarrow}

\newcommand{\sR}{\mathbb R}
\newcommand{\sN}{\mathbb N}
\newcommand{\sC}{\mathbb C}

\newcommand{\pvec}{(p_1, \dots, p_n)}
\newcommand{\qvec}{(q_1, \dots, q_m)}

\newcommand{\escp}[1]{p_{#1}^{(\alpha)}}

\newcommand{\Hg}{\mc H}
\newcommand{\Ht}{\mc T}
\newcommand{\Hsm}{\mc D}

\newcommand{\opa}{\op_\gamma}

\newcommand{\oplam}{\op_\gamma}
\newcommand{\lama}{\lambda}
\newcommand{\lamg}{\gamma}

\begin{document}

\begin{frontmatter}

\title{\Large\bf Generalized Shannon-Khinchin Axioms and \\ Uniqueness Theorem for
Pseudo-additive Entropies }
\tnotetext[t1]{Research supported by Ministry  of Science and
Technological Development, Republic of Serbia, Grants No. 174026
and  III 044006}

\author[misanu]{Velimir M. Ili\'c\corref{cor}}
\ead{velimir.ilic@gmail.com}

\author[fosun]{Miomir S. Stankovi\'c}
\ead{miomir.stankovic@gmail.com}

\cortext[cor]{Corresponding author. Tel.: +38118224492; fax:
+38118533014.}
\address[misanu]{Mathematical Institute of the Serbian Academy of Sciences and Arts, Kneza Mihaila 36, 11000 Beograd, Serbia}
\address[fosun]{University of Ni\v s, Faculty of Occupational Safety, \v Carnojevi\'ca 10a, 18000 Ni\v s, Serbia}

\begin{abstract}
\begin{minipage}{15.5cm}

We consider the Shannon-Khinchin axiomatic systems for the
characterization of generalized entropies such as
Sharma-Mital and Frank-Daffertshofer entropy. We provide the generalization of
Shannon-Khinchin axioms and give the corresponding uniqueness
theorem. The previous attempts at such axiomatizations are also
discussed.
\end{minipage}
\end{abstract}

\end{frontmatter}

\section{Introduction}
\label{sec: intro}

In the past, there was extensive work on defining the information
measures which generalize the Shannon entropy \cite{Khinchin_57}
and on deriving the axiomatic systems for their characterization.
Shannon entropy follows the composition rule by which the entropy
of a joint system can be represented as the sum of the entropy of
one system and the conditional entropy of another, with respect to
the first one, with standardly defined expectation operator. The
composition rule is shown to be of the essential properties of
Shannon entropy and the axiomatic characterization based on this
property has been established in literature as the
Shannon-Khinchin axiomatic system \cite{Khinchin_57}. Since then,
there has been extensive work on generalizing the Shannon entropy
and its axiomatic characterization.

R\'enyi introduced a one-parameter generalization of the Shannon
entropy in the way that expectation operator is defined as
quasi-linear mean \cite{Renyi_07}, and the generalization of
Shannon-Khinchin axioms for the R\'enyi entropy is presented by
Jizba and Arimitsu \cite{Jizba_Arimitsu_04}. Another approach for
the generalization of the Shannon-Khinchin axioms is to keep the
linear expectation form, but to replace the real addition in the
composition rule by a more general one. Such approach was first
proposed by Abe \cite{Abe_97}, who used the so-called $\opa$
addition \cite{Nivanen_Wang__03} and obtained the
Havrda-Charv\'at-Tsallis entropy \cite{Havrda_Charvat_67},
\cite{Tsallis_88} as the unique entropy which sattify his
axiomatic system.

In this paper, we  first relax the so-called normalization axiom
from the Jizba-Arimitsu axiomatic system \cite{Jizba_Arimitsu_04},
and show that such a system characterizes a two-parameter
generalization of Shannon and Re\'nyi entropy previously
introduced by Nath \cite{Nath_68b}. After that, we show that
$\opa$ operation in Abe's axiomatic system \cite{Abe_97} can be
replaced with a more general pseudo-addition operation without
affecting the resulting entropy form. Accordingly, the
Havrda-Charv\'at-Tsallis entropy is a unique entropy which can be
characterized with generalized Shannon-Khinchin axioms if the
linear expectation is used. Finally,
we present the generalized Shannon-Khinchin axiomatic system based
on the combination of the two mentioned approaches. A similar
class of entropies has previously been characterized in
\cite{Amblard_Vignat_06} and \cite{Ilic_et_al_cert_inf} by the set
of axioms which assume that pseudo-additive entropy is represented
as quasi-linear mean-value of pseudo-additive information content
and the statistical properties of such entropies are examined in
\cite{Esteban_Morales_95}. A similar attempt for the
generalization has been made in \cite{Jizba_Arimitsu_04a}, where
the authors use pseudo-linear expectation and $\opa$ addition for
the composition rule, but the class of entropies derived in
\cite{Jizba_Arimitsu_04a} is incomplete, which is also discussed
in the present paper.

The paper is organized as follows. In section \ref{sec: shannon}
we review the Shannon-Khinchin axioms. The generalization of the
Shannon-Khinchin axioms for the Nath and R\'enyi entropy is
presented in section \ref{sec: varma} and for the
Havrda-Charv\'at-Tsallis entropy in section \ref{sec: tsallis}. In
section \ref{sec: gsk axioms} we consider the generalization of
Shannon-Kihinchin axioms for the class of entropies which can be
represented as a nonlinear transformation of the Nath entropy. Two
special cases, Sharma-Mittal and the Frank-Daffertshofer
entropies, are considered in subsection \ref{sec: gsk axioms:
subsec: sm}.

\newcommand{\Hs}{\mc S}
\newcommand{\sRp}{\sR^+}

\section{Shannon-Khinchin axioms}
\label{sec: shannon}

Let the set of all $n$-dimensional distributions be denoted with
\begin{equation}
   \Delta_n \equiv \left\{ (p_1, \dots , p_n) \Big\vert \; p_i \ge 0,
      \sum_{i=1}^{n} p_i = 1 \right \}, \quad n>1.
  \label{Delta}
\end{equation}
and let $\sR^+$ denotes the set of positive real numbers. The Shannon entropy of $n$-dimensional distribution 
is a function $\Hs_n:\Delta_n\ra\sRp$ from the family
parameterized by $c \in \sR$:
\begin{align}
\label{Shannon: H final c, alpha} \Hs_n(\mcP) =
 \cop \cdot \sum_{k=1}^{n} p_k  \log_2 p_k, \quad \cop<0,
\end{align}
%
The following theorem characterizes the Shannon entropy by the so
called Shannon-Khinchin axioms \cite{Khinchin_57}.

\begin{theorem}\rm
\label{Reny 2: theorem}

Let the function $\Hs_n:\Delta_n\ra\sRp$ satisfy the following
Shannon-Khinchin axioms, for all $n\in\sN$, $n>1$:

\begin{description}

\item{[SA1]} $\Hs_n$ is continuous in $\Delta_n$;

\item{[SA2]} $\Hs_n$ takes its largest value for the uniform
distribution, $\mcU_n=\left({1}/{n}, \dots, {1}/{n}\right) \in
\Delta_n$, i.e. 
$\Hs_n(\mcP)\le \Hs_n(\mcU_n)$, for any $\mcP \in \Delta_n$;

\item{[SA3]} $\Hs_n$ is expandable: $\Hs_{n+1}(p_1,p_2, \ldots , p_n, 0 ) = \Hs_n(p_1,p_2, \ldots ,
p_n)$ for all $(p_1,\dots,p_n)\in\Delta_n$;

\item{[SA4]} Let $\mcP = (p_1, \dots, p_n) \in \Delta_n$, 
$\mcPQ = (r_{11}, r_{12}, \dots, r_{nm}) \in \Delta_{nm}$, $n, m
\in \sN$, $n,m>1$ such that $p_i = \sum_{j=1}^m r_{i j}$, and
$\mcQ_{ | k} = (q_{1|k}, \dots, q_{m|k}) \in \Delta_m$, where
$q_{i|k} = r_{ik}/p_k$. Then,
\begin{equation}
\Hs_{nm}(\mcPQ) = \Hs_n(\mcP) + \Hs_m(\mcQ| \mcP),
\quad\text{where}\quad
\Hs_m(\mcQ| \mcP) = \sum_k p_k\cdot \Hs_m(\mcQ_{ | k}).
\end{equation}

\end{description}
Then, the function $\Hs_n$ is the Shannon entropy given by the
class (\ref{Shannon: H final c, alpha}).
\end{theorem}

\section{Generalized Shannon-Khinchin axioms for Nath and R\'enyi entropy}
\label{sec: varma}

The Nath entropy of $n$-dimensional distribution 
is a function $\Hv_n: \Delta_n \ra \sRp$ from the family
parameterized by $\cop,\lambda,\alpha \in \sR$:
\begin{align}
\label{Nath entropy: Nath definition} \Hv_n(\mcP) =
\begin{dcases}
\ \cop \cdot \sum_{k=1}^{n} p_k  \log_2 p_k, \quad\quad\quad \
\cop <0, \quad &\mbox{for} \quad \lambda =0,
\\
\ \frac{1}{\lambda} \log_2 \left( \sum_{k=1}^{n} p_k^{\alpha}
\right), \quad \alpha>0,\ \lambda\cdot(1 - \alpha) > 0, \quad
&\mbox{for} \quad \lambda \neq 0.
\end{dcases}
\end{align}
If $\lambda = 1- \alpha$ and $\cop=-1$, the Nath entropy reduces to the R\'enyi entropy, 
which is a function $\mc R_n: \Delta_n \ra \sRp$,
\begin{align}
\label{Nath entropy: Renyi definition} \mc R_n(\mcP) =
\begin{dcases}
\quad - \sum_{k=1}^{n} p_k  \log_2 (p_k), \quad &\mbox{for} \quad
\alpha =1,
\\
\ \frac{1}{1-\alpha} \log_2 \left( \sum_{k=1}^{n} p_k^{\alpha}
\right),\ \alpha>0, \quad &\mbox{for} \quad \alpha \neq 1.
\end{dcases}
\end{align}

Previously, Jizba and Arimitsu \cite{Jizba_Arimitsu_04} had given
the characterization of R\'enyi entropy using the generalized
Shannon-Khinchin axioms. 
In this section, we extend the results from
\cite{Jizba_Arimitsu_04} to the more general case of the Nath
entropy.

\begin{theorem}\rm
\label{Nath entropy: uniquenes theorem} Let the function
$\Hv_n:\Delta_n\ra\sRp$ satisfy the following generalized
Shannon-Khinchin axioms for all $n\in\sN$, $n>1$:

\begin{description}

\item {[NSK1]} $\Hv_n$ is continuous in $\Delta_n$;

\item{[NSK2]} $\Hv_n$  takes its largest value for the uniform
distribution, $\mcU_n=\left({1}/{n}, \dots, {1}/{n}\right) \in
\Delta_n$, $\Hv_n(\mcP)\le \Hv_n(\mcU_n)$, for any $\mcP \in
\Delta_n$;

\item{[NSK3]} $\Hv_n$ is expandable: $\Hv_{n+1}(p_1,p_2, \ldots , p_n, 0 ) = \Hv_n(p_1,p_2, \ldots ,
p_n)$ for all $(p_1,\dots,p_n)\in\Delta_n$;

\item{[NSK4]} Let $\mcP = (p_1, \dots, p_n) \in \Delta_n$, 
$\mcPQ = (r_{11}, r_{12}, \dots, r_{nm}) \in \Delta_{nm}$, $n, m
\in \sN$, $n,m>1$ such that $p_i = \sum_{j=1}^m r_{i j}$, and
$\mcQ_{ | k} = (q_{1|k}, \dots, q_{m|k}) \in \Delta_m$, where
$q_{i|k} = r_{ik}/p_k$ and $\alpha \in \sRp$ is some fixed
parameter. Then,
\begin{equation}
\Hv_{nm}(\mcPQ) = \Hv_n(\mcP) + \Hv_m(\mcQ| \mcP),
\quad\text{where}\quad
\Hv_m(\mcQ| \mcP) = f^{-1} \left(\sum_{k=1}^n \escp{k}
f(\Hv_m(\mcQ_{ | k})) \right),
\end{equation}
where $f$ is an invertible continuous function and the
\emph{$\alpha$-escort distribution}
$\mcP^{(\alpha)}=(\escp{1},\dots,\escp{n})\in \Delta_n$ of
distribution $\mcP \in \Delta_n$ is defined with
\begin{equation}
\escp{k}=\frac{p_k^\alpha}{\sum_{i=1}^n p_i^\alpha},\quad
k=1,\dots,n, \quad \alpha >0.
\end{equation}

\end{description}

Then, $f$ is a function from the class parameterized by $b, c, d,
\lambda, \gamma \in \sR$:
\begin{equation}
\label{Nath: theorem: f form} f(x) =
\begin{dcases}
\ \ c \cdot x + b,\quad\quad\
\ c\neq 0, &\text{for}\quad\lambda = 0, \\
\ \ \frac{d \cdot 2^{\lambda x}-1}{\gamma},\quad d,\gamma\neq 0,
&\text{for}\quad\lambda \neq 0,
\end{dcases}
\end{equation}
and the function $\Hv_n$ is the Nath entropy given by the class
(\ref{Nath entropy: Nath definition}). In addition, if the
following normalization axiom holds:
\begin{description}
\item{[NSK5]} $\Hv_2\left(\frac{1}{2},\frac{1}{2}\right)=1$,
\end{description}
$\Hv_n$ reduces to the R\'enyi entropy given by the class
(\ref{Nath entropy: Renyi definition}).

\end{theorem}

In the proof of the theorem we will use the following theorem from
\cite{Jizba_Arimitsu_04} and \cite{Ilic_et_al_cert_inf}.

\begin{theorem}\rm
\label{Reny 1: theorem} Let $g:\sR \ra \sR$ be continuous
invertible function and $\Hv_n: \Delta_n \ra \sRp$ is continious
function,
\begin{equation}
\Hv_{n}(\mcP) = g^{-1} \left( \sum_{k=1}^{n} \escp{k} g(\cop\cdot
\log_2 p_k) \right),  \quad \cop <0, \quad \alpha>0,
\end{equation}
for all $n\in\sN$, $n>1$ and $\mcP = (p_1, \dots, p_n) \in
\Delta_n$, and let $\Hv_n$ be additive, i.e.
$\Hv_{nm}(\mcP\star\mcQ)=\Hv_n(\mcP) + \Hv_m(\mcQ)$ for all $\mcP
= (p_1, \dots, p_n) \in \Delta_n$, $\mcQ = (q_1, \dots, q_m) \in
\Delta_m$ $n, m \in \sN$, $n,m>1$. Then, $g$ is the function from
the class parameterized by $c, \lambda, \gamma \in \sR \setminus
\{0\}$:
\begin{equation}
\label{Reny 1: theorem: g(x)} g(x) =
\begin{dcases}
\ \ -c \cdot x, &\text{for}\quad\lambda = 0 \\
\frac{2^{-\lambda x}-1}{\gamma}, &\text{for}\quad\lambda \neq 0
\end{dcases}
\quad\quad\Leftrightarrow\quad\quad g^{-1}(x) =
\begin{dcases}
\ \ -\frac{1}{c}\cdot x, &\text{for}\quad\lambda = 0 \\
-\frac{1}{\lambda} \log_2{\left(\gamma x + 1\right)},
&\text{for}\quad\lambda \neq 0
\end{dcases}
\end{equation}
The entropy is uniquely determined with
\begin{eqnarray}
\label{renyi: H with c} \Hv_n(\mcP) =
\begin{dcases}
\cop \cdot \sum_{k=1}^{n} \escp{k} \log_2 p_k, &\text{for}\quad \lambda=0 \\
-\dfrac{1}{\lambda}%
\log_2 \left( \dfrac{\sum_{k=1}^n p_k^{\alpha-\cop
\lambda}}{\sum_{k=1}^n p_k^{\alpha}}\right), &\text{for}\quad
\lambda\neq 0,
\end{dcases}
\label{rensh1}
\end{eqnarray}
where $\cop<0$, $\alpha>0$ and $\alpha-\cop \lambda > 0$.
\end{theorem}

\textbf{Proof of the Theorem \ref{Nath entropy: uniquenes
theorem}.}
Let $\mc L(r) = \Hv_r(\mc U_r)$ denote the entropy of the uniform
distribution $\mcU_r=\left({1}/{r}, \dots, {1}/{r}\right) \in
\Delta_r$. By successive usage of axioms [NSK2] and [NSK3] we
conclude that $\mc L(r) \le \mc L(r+1)$, i.e. $\mc L$ is a
non--decreasing function.
Let for $\mcP = \pvec \in \Delta_n$ and $\mcQ = \qvec \in
\Delta_m$ the \textit{direct product}, $\mcP \star \mcQ \in
\Delta_{nm}$, be defined as
\begin{equation}
\mcP \star \mcQ = (p_1q_1, p_1q_2, \dots, p_n q_m).
\end{equation}
Repeated application of axiom [NSK4] then leads to
\begin{eqnarray}
\label{Renyi 2: Khinchin equation} {\Hv_{r^m}}(\underbrace{\mc U_r
\star \mc U_r \star \ldots \star \mc U_r}_{m \text{ times}}) =
\sum_{k=1}^m \Hv_r(\mc U_r) = m \cdot \Hv_r(\mc U_r) \quad
\hbox{i.e.} \quad {\mathcal{L}}(r^m)= m \cdot \mathcal{L}(r).
\label{VI1}
\end{eqnarray}
Since $\mc L$ is non--decreasing, the equation (\ref{Renyi 2:
Khinchin equation}) has unique solution \cite{Khinchin_57}
\begin{equation}
\mc L(r) = -\cop \cdot \log_2 r, \quad \cop<0.
\end{equation}

Let us now determine the entropy form for the distribution $\mcP =
(p_1, \dots, p_n) \in \Delta_n$ when $p_i$ are rational numbers
and the case for irrational numbers follows from the continuity of
entropy. Let  $\mcP = (p_1, \dots, p_n) \in \Delta_n$ , $\mcQ_{|k}
= (q_{1|k}, \dots, q_{m_k|k}) \in \Delta_{m_k}; k=1,\dots,n$ and
$\mcPQ = (r_{11}, r_{12}, \dots, r_{nm}) \in \sC^{nm}$, for $n, m
\in \sN$, $n,m>1$ and %
$p_i=m_i/m$; $\quad r_{ij}=1/m$, $q_{j|i}=1/m_i$,
where $m=\sum_{i=1}^n m_i$ and $m_i \in \sN$, $m_i>1$ for any $i =
1, \dots, n$ and $j = 1,\dots,m_i$. Then we have
$\Hv_{m}(\mcP\mcQ)=\Hv_{m}(\mcU_{m})=\mc L(m) = -\cop \cdot \log_2
m$ and $\Hv_{m_k}(\mcQ_{|k})=\Hv_{m_k}(\mcU_{m_k})=\mc L(m_k)=
-\cop \cdot \log_2 m_k$. Since $p_i = \sum_{j=1}^m r_{i j}$, and
$q_{i|k} = r_{ik}/p_k$, we can apply the axiom [{NSK4}] to obtain
\begin{equation}
\label{Reny 1: theorem: ra2 for rational} \Hv_n(\mcP) =
-\cop \cdot \log_2 m -%
f^{-1}\left(\sum_{k=1}^n \escp{k} f\left(-\cop \cdot \log_2
m_k\right)\right) = -\cop \cdot \log_2 m -f^{-1} \left(
\sum_{k=1}^n \escp{k} f\left(-\cop \cdot \log_2 p_k - \cop \cdot
\log _2 m \right)\right)\, .
\end{equation}
Let us define $f_y(x) = f(-x-y)$, and $f^{-1}_y(z) =
-y-f^{-1}(z)$. If we set $y=\cop\cdot\log_2 m$ the equality
(\ref{Reny 1: theorem: ra2 for rational}) becomes
\begin{equation}
\Hv_n(\mcP) = f^{-1}_{y} \left( \sum_{k=1}^{n}
\escp{k} f_{y}\left(\cop\cdot\log_2 p_k \right) \right),\quad\quad 
\cop<0. \label{Nath 2: H L(g)}
\end{equation}
Since $f$ is continuous, both $f_y$ and $f_y^{-1}$ are continuous,
as well as the entropy, and we may extend the result (\ref{Nath 2:
H L(g)}) from rational $p_k$'s to any real valued $p_k$'s defined
in [0,1].
Now, if we the axiom [{NSK4}] is used with $\mcPQ = \mcP \star
\mcQ$, the conditions from the Theorem \ref{Reny 1: theorem} are
satisfied so that the function $f_y$ is uniquely determined by the
class (\ref{Reny 1: theorem: g(x)}):
\begin{equation}
\label{Reny 1: theorem: fy(x)} f_y(x) =
\begin{dcases}
\ \ -c \cdot x, &\text{for}\quad\lambda = 0, \\
\frac{2^{-\lambda x}-1}{\gamma}, &\text{for}\quad\lambda \neq 0,
\end{dcases}
\end{equation}
where $c, \lambda, \gamma \in \sR \setminus \{0\}$. The entropy is
uniquely determined by the class (\ref{renyi: H with c})
parameterized by $\cop$, $\alpha$ and $\lambda$. The relationship between the parameters 
is determined by usage of the axiom [{NSK4}], in the same manner
as in \cite{Ilic_et_al_Nath_analicity} and
\cite{Jizba_Arimitsu_04}, which gives $\alpha=1$ for $\lambda=0$
and  $\alpha - \cop\lambda=1$ for $\lambda\neq0$. Positivity of
entropy implies $\cop<0$ and $\lambda\cdot(1-\alpha) > 0$. Thus,
if a function satisfies [{NSK1}]-[{NSK4}], then it has the form
(\ref{Nath entropy: Nath definition}). Conversely, if a function
has the form (\ref{Nath entropy: Nath definition}), [{NSK1}],
[{NSK3}] and [{NSK4}] are obviously satisfied. In addition the
R\'enyi entropy attains maximum for uniform distribution, as shown
in \cite{Jizba_Arimitsu_04}, so that [{NSK4}] is satisfied, since
$\Hv_n(\mcP)=\frac{1 - \alpha}{\lambda} \cdot \mc R_n(\mcP)$ and
${\lambda\cdot(1 - \alpha)}>0$.

The form of function $f$ can be determined by substitution of
(\ref{Nath entropy: Nath definition}) in [{NSK4}]. Let $\mcP = (p_1, \dots, p_n) \in \Delta_n$, 
$\mcPQ = (r_{11}, r_{12}, \dots, r_{nm}) \in \Delta_{nm}$, $n, m
\in \sN$, $n,m>1$ such that $p_i = \sum_{j=1}^m r_{i j}$, and
$\mcQ_{ | k} = (q_{1|k}, \dots, q_{m|k}) \in \Delta_m$, where
$q_{i|k} = r_{ik}/p_k$. If we introduce $\tilde f(x)=2^{\lambda
x}$ for $\lambda\neq 0$ and $\tilde f(x)=x$ for $\lambda=0$
(recall that in this case $\alpha=1$), it is easy to obtain
\begin{equation}
\Hv_{nm}(\mcP\star\mcQ)-\Hv_n(\mcP)=%
{\widetilde f}^{-1} \left(\sum_{k=1}^n \escp{k} {\widetilde
f}(\Hv_m(\mcQ_{ |
k})) \right)= %
f^{-1} \left(\sum_{k=1}^n \escp{k} f(\Hv_m(\mcQ_{ | k})) \right).
\end{equation}
Accordingly, $f$ and $\widetilde f$ generate the same mean so
that, as shown in \cite{ Hardy_et_al_34}, $f$ is a linear function
of $\widetilde f$ and the form (\ref{Nath: theorem: f form})
follows (invertibility of implies $c,d\neq=0$.

Finally, if in addition [{NSK5}] holds, from (\ref{Nath entropy:
Nath definition}) we get $\lambda = \alpha-1$ and $c=-1$ and the
theorem is proven.

\section{Generalized Shannon-Khinchin axioms for Havrda-Charv\'at-Tsallis entropy}
\label{sec: tsallis}

Let $\isoRtoX: \sR \ra \sR$ be an increasing continuous (hence
invertible) function such that $h(0)=0$ and let the
pseudo-addition operation $\op$ be defined as:
\begin{equation}
\label{sec: tsallis: op definition} \isoRtoX(x+y) = \isoRtoX(x)
\op \isoRtoX(y); \quad x, y \in \sR. 
\end{equation}
If the mapping $h$ is parameterized by $a, \lama, \lamg \in \sR$ and defined
with
\begin{equation}
\label{sec: tsallis: h}
\isoRtoX(x) = %
\begin{dcases}
\ a \cdot x, \quad \quad\quad a > 0, \quad &\mbox{ for }\quad \lambda = 0, \\
\frac{2^{\lama \cdot x} - 1}{\lamg}, \quad \lama\cdot\lamg> 0, \quad &\mbox{ for }\quad \lambda
\neq 0,
\end{dcases}
\end{equation}
the formula (\ref{sec: tsallis: op definition}),
defines $\oplam$-addition \cite{Nivanen_Wang__03},
\begin{equation}
\label{SM: op} u \oplam v = u + v +\lamg u v; \quad u, v \in \sR.
\end{equation}
For the case $\lamg=0$, $h$ reduces to a linear function and the
$\oplam$-addition reduces to ordinary addition.

In \cite{Abe_00}, Abe considers generalized Shannon-Khinchin
axiomatic system in which the first three axioms are the same as
[SA1]-[SA3], but in the fourth axiom, $\oplam$-addition is used
instead of the ordinary one and the expectation is taken with
respect to the $\alpha$-escort distribution. More specifically,
Abe's fourth axiom is:

\begin{description}

\item{[ASK4]} Let $\mcP = (p_1, \dots, p_n) \in \Delta_n$, 
$\mcPQ = (r_{11}, r_{12}, \dots, r_{nm}) \in \Delta_{nm}$, $n, m
\in \sN$ $,n,m>1$, such that $p_i = \sum_{j=1}^m r_{i j}$, and
$\mcQ_{ | k} = (q_{1|k}, \dots, q_{m|k}) \in \Delta_m$, where
$q_{i|k} = r_{ik}/p_k$. Then,
\begin{equation}
\Ht_{nm}(\mcPQ) = \Ht_n(\mcP)
\oplam \Ht_m(\mcQ| \mcP),
\quad\text{where}\quad \Ht_m(\mcQ| \mcP) = \sum_{k=1}^n \escp{k}
\Ht_m(\mcQ_{ | k})), \quad \alpha>0.
\end{equation}
\end{description}
In \cite{Abe_00}, it is shown that the axiomatic system
[SA1]-[SA3] and [ASK4], uniquely characterizes the
Havrda-Charv\'at \cite{Havrda_Charvat_67} and Tsallis
\cite{Tsallis_88} entropies\footnote{Actually, Abe consider the
case of Tsallis entropy only, which is obtained for
$\gamma=1-\alpha$, and does not cover the case of Havrda-Charv\'at
entropy obtained for $\gamma = 2^{1-\alpha}-1$. However, the
discussion holds in our general case, which can be
straightforwardly shown by repeating the steps from
\cite{Abe_00}}:
\begin{equation}
\label{tsallis entropy: definition} \Ht_n(\mcP) =
\begin{dcases}
\mc S_n(\mcP) = \cop \cdot \sum_{k=1}^{n} p_k  \log_2 p_k, \quad \cop<0,
&\mbox{ for }\quad \lamg = 0\\
\quad \frac{1}{\gamma} \cdot \left(\sum_k p_k^\alpha - 1\right), \
\quad \alpha>0,\ \lamg(1-\alpha)>0, &\mbox{ for }\quad \lamg \neq 0
\end{dcases}
\end{equation}
In the following theorem we show that the entropy form
(\ref{tsallis entropy: definition}) can be characterized with
weaker assumption about the pseudo-addition defined with
(\ref{sec: tsallis: op definition}). Alternatively,
$\oplam$-addition is the unique pseudoaddition operation which can
be used in axiom [ASK4].

\begin{theorem}\rm
Let the entropy $\Ht_n: \Delta_n \ra \sRp$ be defined as a
function which for all $n\in\sN$, $n>1$ satisfies the following
axioms:

\begin{description}

\item {[ASK1]} $\Ht_n$ is continuous in $\Delta_n$;

\item{[ASK2]} $\Ht_n$ takes its largest value for the uniform
distribution, i.e. for any $\mcP \in \Delta_n$, $\Ht(\mcP)\le
\Ht_n(\mcU_n)$;

\item{[ASK3]} $\Ht_n$ is expandable: $\Ht_{n+1}(p_1,p_2, \ldots , p_n, 0 ) = \Ht_n(p_1,p_2, \ldots ,
p_n)$ for all $(p_1,\dots,p_n)\in\Delta_n$;

\item{[ASK4]} Let $\mcP = (p_1, \dots, p_n) \in \Delta_n$, 
$\mcPQ = (r_{11}, r_{12}, \dots, r_{nm}) \in \Delta_{nm}$, $n, m
\in \sN$, $n,m>1$ such that $p_i = \sum_{j=1}^m r_{i j}$, and
$\mcQ_{ | k} = (q_{1|k}, \dots, q_{m|k}) \in \Delta_m$, where
$q_{i|k} = r_{ik}/p_k$. Then,
\begin{equation}
\label{tsallis pseudoadditivity} \Ht_{nm}(\mcPQ) = \Ht_n(\mcP) \op
\Ht_m(\mcQ| \mcP),
\quad\text{where}\quad \Hg_m(\mcQ| \mcP) = \sum_k \escp{k}
\Hg_m(\mcQ_{ | k})), \quad \alpha>0,
\end{equation}
where $\op$-addition is defined with (\ref{sec: tsallis: op
definition}).

\end{description}

Than, the entropy has the form (\ref{tsallis entropy:
definition}).

\end{theorem}

\textbf{Proof.} Let $h^{-1}$ denote the inverse mapping of $h$,
let us denote $\Hv_n(\mcP)=h^{-1}(\Ht_n(\mcP)) \Leftrightarrow
\Ht_n(\mcP)=h(\Hv_n(\mcP))$ and let us apply $h^{-1}$ to,
[ASK1]-[ASK4]. Since $h^{-1}$ is increasing and continuous
[ASK1]-[ASK4] transforms into [NSK1]-[NSK4] (with $f \equiv h$)and
we get
\begin{align}
\Hv_n(\mcP) =
\begin{dcases}
\ \cop \cdot \sum_{k=1}^{n} p_k  \log_2 (p_k) ; \quad\quad\quad \
\cop <0 \quad &\mbox{for} \quad \lambda =1,
\\
\ \frac{1}{\lambda} \log_2 \left( \sum_{k=1}^{n} p_k^{\alpha}
\right); \quad \alpha>0,\ \lambda(1 - \alpha) > 0\  \quad &\mbox{for} \quad \lambda \neq 0.
\end{dcases}
\end{align}
The function $h$ is given with (\ref{Nath: theorem: f form}), and
using $h(0)=0$ we get (\ref{sec: tsallis: h}):
\begin{equation}
\isoRtoX(x) = %
\begin{dcases}
\ a \cdot x, \quad \quad\quad a > 0, \quad &\mbox{ for }\quad \lambda = 0, \\
\frac{2^{\lama \cdot x} - 1}{\lamg}, \quad \lama\cdot\lamg> 0, \quad &\mbox{ for }\quad \lambda
\neq 0,
\end{dcases}
\end{equation}
and since $\Ht_n(\mcP)=h(\Hv_n(\mcP))$, the theorem is proven.

\section{Generalized Shannon-Khinchin Axioms for a nonlinear
transformation of the Nath entropy}

\label{sec: gsk axioms}

In this section we combine the axiomatic systems from sections
\ref{sec: varma} and \ref{sec: tsallis}. Thus, we consider the
pseudo-addition generated by an increasing continuous function
$\isoRtoX: \sR \ra \sR$ such that $h(0)=0$ by the following
equation:
\begin{equation}
\label{sec: gen entr: op definition} \isoRtoX(x + y) = \isoRtoX(x)
\op \isoRtoX(y); \quad x,y \in \sR,
\end{equation}
and we characterizes the entropy 
$\Hg_n: \Delta_n \ra \sRp$, as a function which for all $n\in\sN$,
$n>1$ satisfies the following axioms:

\begin{description}

\item {[GSK1]} $\Hg_n$ is continuous in $\Delta_n$;

\item{[GSK2]} $\Hg_n$ takes its largest value for the uniform
distribution, i.e. for any $\mcP \in \Delta_n$, $\Hg(\mcP)\le
\Hg_n(\mcU_n)$;

\item{[GSK3]} $\Hg_n$ is expandable: $\Hg_{n+1}(p_1,p_2, \ldots , p_n, 0 ) = \Hg_{n}(p_1,p_2, \ldots ,
p_n)$ for all $(p_1,\dots,p_n)\in\Delta_n$;

\item{[GSK4]} Let $\mcP = (p_1, \dots, p_n) \in \Delta_n$, 
$\mcPQ = (r_{11}, r_{12}, \dots, r_{nm}) \in \Delta_{nm}$, $n, m
\in \sN$, $n,m>1$ such that $p_i = \sum_{j=1}^m r_{i j}$, and
$\mcQ_{ | k} = (q_{1|k}, \dots, q_{m|k}) \in \Delta_m$, where
$q_{i|k} = r_{ik}/p_k$ and $\alpha \in \sRp$ is some fixed
parameter. Then,
\begin{equation}
\Hg_{nm}(\mcPQ) = \Hg_n(\mcP) \op \Hg_m(\mcQ| \mcP),
\quad\text{where}\quad
\Hg_m(\mcQ| \mcP) = g^{-1} \left(\sum_k \escp{k} g(\Hg_m(\mcQ_{ |
k})) \right),
\end{equation}
where $g$ is the invertible continuous function.

\end{description}

In the following theorem we establish the class of entropies
characterized by the generalized Shannon-Khincin axioms.

\begin{theorem}\rm
\label{sec: gen entr: theorem} A function $\Hg_n: \Delta_n \ra
\sRp$,  satisfies generalized Shannon-Khinchin axioms
[GSK1]-[GSK4] iff it belongs to the following family parameterized
by $\cop, \lambda, \alpha \in \sR$:
\begin{align}
\Hg_n(\mcP) =\isoRtoX \left(\Hv_n(\mcP)\right)
\begin{dcases}
\ \isoRtoX \left(\cop \cdot \sum_{k=1}^{n} p_k  \log_2 p_k\right),
\quad\quad\quad \ \cop <0, \quad &\mbox{for} \quad \lambda =0,
\\
\ \isoRtoX \left(\frac{1}{\lambda} \log_2 \left( \sum_{k=1}^{n}
p_k^{\alpha} \right)\right), \quad \alpha>0,\ \lambda\cdot(1 -
\alpha) > 0, \quad &\mbox{for} \quad \lambda \neq 0.
\end{dcases}
\end{align}

In addition, if the following normalization axiom
holds:
\begin{description}
\item{[GSK5]}  Normalization axiom: $\Hg_2(\frac{1}{2},\frac{1}{2}) = \isoRtoX(1)$,
\end{description}
then $\cop = -1$, $\lambda = 1-\alpha$ and $\Hg_n$ belongs to the
family
\begin{equation}
\label{Renyi 2: H final}%
\Hg_n(\mcP)=  \isoRtoX \left( \mc R_n(\mcP) \right) =
\begin{dcases}
\isoRtoX \left( - \sum_{k=1}^{n} p_k \log_2
(p_k) \right) \quad  &\mbox{for} \quad\quad \alpha =1,\\
\isoRtoX \left( \frac{1}{1 - \alpha} \log_2 \left( \sum_{k=1}^{n}
p_k^{\alpha} \right) \right), \quad &\mbox{for} \quad \alpha>0, \
\alpha \neq 1.
\end{dcases}
\end{equation}

\end{theorem}
\textbf{Proof:} Let $h^{-1}$ denotes the inverse mapping of $h$,
let us denote $\Hv_n(\mcP)=h^{-1}(\Hg_n(\mcP)) \Leftrightarrow
\Hg_n(\mcP)=h(\Hv_n(\mcP))$ and $f=g \comp h$, where $\comp$
denote the composition of functions. Than, [{GSK1}]-[GSK5] reduces
to [{NSK1}]-[{NSK5}] and result follows.

\subsection{Example - Sharma-Mittal and Frank-Daffertshofer entropies} \label{sec: examples: q
addition} \label{sec: gsk axioms: subsec: sm}

The idea for combining axiomatic systems [NSK1]-[NSK5] and
[ASK1]-[ASK5] has firstly been proposed in
\cite{Jizba_Arimitsu_04a} for the case of $\opa$-addition and the
function $h$ given by (\ref{sec: tsallis: h}), which is the
special case of the system [GSK1]-[GSK5] proposed in this section.
If the function (\ref{sec: tsallis: h}) is used and $a=1$,
$\lambda=1-q$, where $q\in \sR$,
\begin{equation}
\label{sec: sm: h}
\isoRtoX(x) = %
\begin{dcases}
\quad\quad x &\mbox{ for }\quad q = 1, \\
\frac{2^{(1-q) \cdot x} - 1}{\lamg}, \quad \gamma\cdot(1-q)> 0,
\quad &\mbox{ for }\quad q \neq 1,
\end{dcases}
\end{equation}
generalized entropy form (\ref{Renyi 2: H final}) reduces to the
two-parameter entropy:

\begin{equation}
\label{sm entropy: definition} \Hsm_n(\mcP) =
\begin{dcases}
\mc S_n(\mcP) = - \sum_{k=1}^{n} p_k  \log_2 (p_k)
&\mbox{ for }\quad q = 1, \alpha = 1,\\
%
\mc{G}_n(\mcP) = \frac{1}{\gamma} \cdot \left(
\prod_k p_k^{(q-1)\cdot p_k}-1 \right), \ \gamma\cdot(1-q)>0,&\mbox{ for }\quad q \neq 1, \alpha = 1, \\
%
\mc{R}_n(\mcP) = \frac{1}{1 - \alpha}\cdot \log_2 \left(
\sum_{k=1}^{n} p_k^{\alpha} \right) &\mbox{ for }\quad
q = 1, \alpha \neq 1, \alpha>0,  \\
%
\quad\quad \frac{1}{\gamma} \cdot \left(\left[ \sum_k p_k^\alpha
\right]^{\frac{q - 1}{\alpha - 1}}-1\right),\quad\quad
\gamma\cdot(1-q)>0,&\mbox{ for }\quad q \neq 1, \alpha \neq 1,
\alpha>0,
\end{dcases}
\end{equation}
considered by Sharma and Mittal \cite{Sharma_Mittal_75}, for
$\gamma=2^{1-q}-1$, and Frank and Daffertshofer \cite{Frank_00},
for $\gamma=1-q$.
The special cases contained in (\ref{sm entropy: definition}) are
the Shannon entropy, $\Hs_n$, \cite{Khinchin_57}, Gaussian
entropy, $\mc G_n$, \cite{Frank_00} and Renyi entropy, $\mc R_n$,
\cite{Renyi_07}. The uniqueness theorem for $\opa$-addition
generalized Shannon-Khinchin axioms straightforwardly follows from
the theorem \ref{sec: gen entr: theorem}.

\begin{theorem}\rm

Let the entropy $\Hsm_n: \Delta_n \ra \sRp$ be defined as a
function which for all $n\in\sN$, $n>1$ satisfies the following
axioms:

\begin{description}

\item {[SM1]} $\Hsm_n$ is continuous in $\Delta_n$;

\item{[SM2]} $\Hsm_n$ takes its largest value for the uniform
distribution, i.e. for any $\mcP \in \Delta_n$, $\Hsm_n(\mcP)\le
\Hsm_n(\mcU_n)$;

\item{[SM3]} $\Hsm_n$ is expandable: $\Hsm_{n+1}(p_1,p_2, \ldots , p_n, 0 ) = \Hsm_{n}(p_1,p_2, \ldots ,
p_n)$ for all $(p_1,\dots,p_n)\in\Delta_n$;

\item{[SM4]} Let $\mcP = (p_1, \dots, p_n) \in \Delta_n$, 
$\mcPQ = (r_{11}, r_{12}, \dots, r_{nm}) \in \Delta_{nm}$, $n, m
\in \sN$, $n>1$ such that $p_i = \sum_{j=1}^m r_{i j}$, and
$\mcQ_{ | k} = (q_{1|k}, \dots, q_{m|k}) \in \Delta_m$, where
$q_{i|k} = r_{ik}/p_k$ and $\alpha \in \sRp$ is some fixed
parameter. Then,
\begin{equation}
\Hsm_{nm}(\mcPQ) = \Hsm_n(\mcP) \oplam \Hsm_m(\mcQ| \mcP),
\quad\text{where}\quad \Hsm_m(\mcQ| \mcP) = f^{-1} \left(\sum_k
\escp{k} f(\Hsm_m(\mcQ_{ | k})) \right),
\end{equation}
where $f$ is invertible continuous function.

\item{[SM5]}  Normalization axiom:
\begin{equation}
\Hsm_2\left(\frac{1}{2},\frac{1}{2}\right) =
\isoRtoX(1) = %
\begin{dcases}
\quad\ x, &\mbox{ for }\quad q = 1, \\
\frac{2^{1-q} - 1}{\lamg}, \quad \gamma\cdot(1-q)> 0, \quad
&\mbox{ for }\quad q \neq 1.
\end{dcases}
\end{equation}

\end{description}

Then, the entropy has the form (\ref{sm entropy: definition}).

\end{theorem}

\begin{remark}\rm
The axiomatic system [SM1]-[SM5] has firstly been proposed in
\cite{Jizba_Arimitsu_04a}. However, the class they obtained is for
$\alpha=1$ solution only. The proof from \cite{Jizba_Arimitsu_04a}
may be rederived by applying the mapping (\ref{sec: tsallis: h})
on [NSK1]-[NSK5] and by setting $\Hsm_n=h \circ \Hv_n$, where $h$
is given with (\ref{sec: sm: h}). The equality (\ref{Nath 2: H
L(g)}) can equivalently be written as
\begin{equation}
\Hsm_n = g^{-1}_{y} \left( \sum_{k=1}^{n} \escp{k} g_{y}(\Logp
p_k^c) \right),
\end{equation}
where $\Logp = \isoXtoR \comp \log_2$, $g_y=f_y \comp \isoXtoR$
and $f_y$ is given with (\ref{Reny 1: theorem: fy(x)}). At this
point Jizba and Arimitsu assume that
\begin{equation}
g^{-1}_{y} \left( \sum_{k=1}^{n} \escp{k} g_{y}(\Logp
p_k^c)\right)= g^{-1} \left( \sum_{k=1}^{n} \escp{k} g(\Logp
p_k^c) \right),
\end{equation}
where $g = f \comp \isoXtoR$ and $f$ is given with (\ref{Nath:
theorem: f form}). However, this implies that $g= f \comp
\isoXtoR$ and $g_y= f_y \comp \isoXtoR$ generate the same mean
and, according to \cite{ Hardy_et_al_34}, $g$ must be a linear
function of $g_y$, which is the case only for
$\lambda=1-\alpha=0$.
\end{remark}

\section{Conclusion}

In this paper we considered a generalization of the
Shannon-Khinchin axiomatic system \cite{Khinchin_57}. Previously,
Jizba and Arimitsu provided the generalization of the
Shannon-Khinchin axioms for the characterization of the R\'enyi
entropy \cite{Jizba_Arimitsu_04}. We modified Jizba-Arimitsu's
system by relaxing the normalization axiom and obtained the Nath
entropy as the unique solution. On the other hand, Abe provided
the axiomatic system for the characterization of the Tsallis
entropy \cite{Abe_00}. Abe's axiomatic system is based on $\opa$
addition \cite{Nivanen_Wang__03}. In this paper we generalized
Abe's axiomatic system by considering the more general
pseudo-addition operation.

In addition, two approaches were combined and the corresponding
uniqueness theorem was given. We obtained a generalized entropy
which can be represented as the nonlinear transformation of the
Nath entropy, whose special case is the Sharma-Mittal entropy
\cite{Sharma_Mittal_75}. Previously, a similar axiomatic system
was discussed in \cite{Jizba_Arimitsu_04a}, and the we commented
on the uniqueness theorem from \cite{Jizba_Arimitsu_04a}.

\bibliographystyle{plain}
\bibliography{reference}

\end{document}